\documentclass[letterpaper, 10 pt, conference]{ieeeconf}  

\IEEEoverridecommandlockouts                              

\usepackage{mathptmx} 
\usepackage{amsmath} 
\usepackage{amssymb}  
\usepackage{flushend}

\usepackage{tikz} 
\usetikzlibrary{positioning, decorations.markings}
\usepackage{forest} 
\usepackage{wrapfig} 
\usepackage{hyperref} 
\usepackage[flushmargin,hang]{footmisc} 
\usepackage{cite} 

\usepackage{amsthm} 

\newtheorem{theorem}{Theorem}
\newtheorem{definition}{Definition}
\newtheorem{lemma}{Lemma}
\newcommand{\bt}{\mathcal{T}}

\newcommand\blue[1]{{#1}}

\title{\LARGE \bf
Continuous-Time Behavior Trees as Discontinuous Dynamical Systems
}

\author{Christopher Iliffe Sprague and Petter \"Ogren
\thanks{Christopher Iliffe Sprague and Petter \"Ogren are with the Robotics, Perception and Learning Lab., School of Electrical Engineering and Computer Science, 
        Royal Institute of Technology (KTH), SE-100 44 Stockholm, Sweden, {\tt\small sprague@kth.se}}%
}

\begin{document}

\maketitle
\thispagestyle{empty}
\pagestyle{empty}

\begin{abstract}
        Behavior trees represent a hierarchical and modular way of combining several low-level control policies into a high-level task-switching policy. Hybrid dynamical systems can also be seen in terms of task switching between different policies, and therefore several comparisons between behavior trees and hybrid dynamical systems have been made, but only informally, and only in discrete time. A formal continuous-time formulation of behavior trees has been lacking. Additionally, convergence analyses of specific classes of behavior tree designs have been made, but not for general designs.
    
        In this letter, we provide the first continuous-time formulation of behavior trees, show that they can be seen as discontinuous dynamical systems (a subclass of hybrid dynamical systems), which enables the application of existence and uniqueness results to behavior trees, and finally, provide  sufficient conditions under which such systems will converge to a desired region of the state space for general designs. With these results, a large body of results on continuous-time dynamical systems can be brought to use when designing behavior tree controllers.
\end{abstract}

\section{Introduction}

Behavior trees \blue{(BTs)} are a way to combine a set of controllers (policies) into higher-level controllers in a hierarchical and modular way. 
In this paper, we give the first continuous-time representation of BTs and provide sufficient conditions for convergence of general BTs.

Modularity is a key tool to handle complexity in software systems, as it enables different components to be developed and tested individually, and BTs have been shown to be optimally modular in comparison to other decision structures \cite{biggar2020modularity}.
Hierarchical modularity, where each module may contain submodules, is also beneficial since a single level of modules in a large system either leads to very large and complex modules, or a very large number of smaller modules. 
Additionally, a hierarchical structure is more natural in many applications, as many tasks can be divided into subtasks in a hierarchical way, such as when a robot has to fetch an object, which might include subtasks such as navigation, door opening, object grasping, and so on.

Improved modularity is the reason that BTs were conceived in the first place \cite{islahandling2005} as an equally expressive \cite{biggar2021expressiveness} alternative to finite-state machines (FSMs) in the design of non-player characters in video games.
In this virtual setting, the world is predictable by design and many low-level policies can be developed with relative ease.
Thus, game developers started to put together large sets of low-level policies earlier than robot developers and therefore had a stronger need for modular tools.
However, the interest in BTs from the robotics community has increased over time and they are now used in both open-source middleware, such as the Robotic Operating System (ROS)\footnote{\url{https://navigation.ros.org/configuration/packages/configuring-bt-navigator.html}} and innovative industry software from Boston Dynamics\footnote{\url{https://dev.bostondynamics.com/docs/concepts/autonomy/missions_service}} and Nvidia\footnote{\url{https://docs.nvidia.com/isaac/isaac/packages/behavior_tree/doc/behavior_trees.html}}.

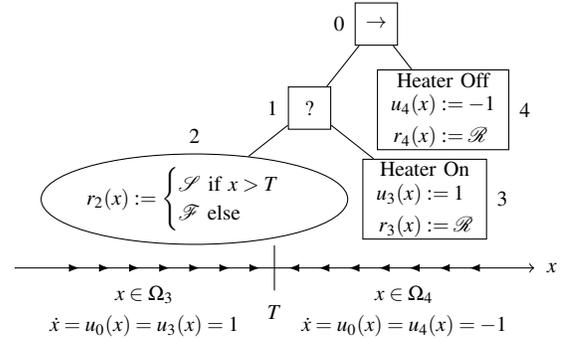
\begin{figure}
        \centering
            \footnotesize
                \begin{forest}
                    for tree={
                        minimum height=2em,
                        minimum width=2em,
                        inner sep=0.1pt,
                    }
                    [$\rightarrow$, draw, tikz={\node[left=1pt of .west]  {$0$};},
                      [$?$, draw, tikz={\node[left=1pt of .west]  {$1$};},
                        [{$
                          r_2(x) := 
                          \begin{cases}
                          \mathcal{S} \text{ if } x > T \\
                          \mathcal{F} \text{ else}
                          \end{cases}
                      $}, draw, ellipse, tikz={\node[above=1pt of .north]  {$2$};}],
                        [{$
                      \begin{array}{c}
                      \text{Heater On} \\
                      \begin{aligned}
                          u_3(x) &:= 1 \\
                          r_3(x) &:= \mathcal{R}
                      \end{aligned}
                      \end{array}
                      $}, draw, tikz={\node[right=1pt of .east]  {$3$};},]
                      ],
                      [{$
                      \begin{array}{c}
                      \text{Heater Off} \\
                      \begin{aligned}
                          u_4(x) &:= -1 \\
                          r_4(x) &:= \mathcal{R}
                      \end{aligned}
                      \end{array}
                      $}, draw, tikz={\node[right=1pt of .east]  {$4$};},]
                    ]
                \end{forest}
        \begin{tikzpicture}
        [decoration={markings, 
            mark= between positions 0.25 and 1 step 4mm 
                with {\arrow{latex}}}
        ] 
        \draw [postaction={decorate}] (-0.4\linewidth,0) -- node[below=1mm] {$\begin{gathered}x \in \Omega_3 \\ \dot{x} = u_0(x) = u_3(x) = 1\end{gathered}$} (0,0); 
        \draw [postaction={decorate}, <-] (0.4\linewidth,0) -- node[below=1mm] {$\begin{gathered}x \in \Omega_4 \\ \dot{x} = u_0(x) = u_4(x) = -1\end{gathered}$} node[right=18mm] {$x$} (0,0); 
        \draw [] (0,-0.3) -- node[below=4mm] {$T$} (0,0.3);
        \end{tikzpicture}
                \caption{A thermostat state-feedback controller modeled by a BT (top), 
            and the phase portrait of its corresponding discontinuous dynamical system $\dot{x} = f(x, u_0(x)) = u_0(x)$ (bottom).
            If $x>T$ then $x\in \Omega_4$ and $\dot x=u_4(x)=-1$. Conversely, if 
            $x\leq T$ then $x\in \Omega_3$ and $\dot x=u_3(x)=1$, see Theorem~\ref{thm:bt_dds}.
            }
            \label{fig:heater_bt}
        \end{figure}
        
\blue{
Even though there is an increasing interest in BTs from the robotics and AI communities (see the recent survey in \cite{iovino2020survey} with over 180 papers) there is still no continuous-time formulation available.
The need for such a formulation is clear from the fact that almost all major branches of control theory, from linear systems to optimal control, have been developed for both continuous-time and discrete-time systems, but 
BTs have so far only had a discrete-time formulation. With the proposed continuous-time model, continuous-time control theory results, such as sliding mode control, can now be used to analyze BT designs.
To date, the only efforts towards continuous-time models have either been informal comparisons of BTs and 
hybrid dynamical systems (HDS), considering discrete-time BTs and discrete-time HDS, or different ways of doing event-based ticking, or letting the tick frequency go to infinity \cite{ogren2012increasing, marzinotto2014towards, colledanchise2016behavior,klockner2014modelica}.

A key topic in control theory is stability and convergence to a particular equilibrium point, or region of the state space.
For a BT, this translates to reaching the so-called success region, a state where the BT returns success. 
Important results on sufficient conditions for convergence to the success region have been presented in 
\cite{paxton2019representing,ogren2020convergence}, but in both cases the analysis was limited to a particular subclass of BTs.
In this letter we propose sufficient conditions that can be use to analyze any BT design.
}

The main contributions of this letter are as follows. We provide the first
formal formulation of BTs in continuous time (Definition \ref{def:behavior}). We
show that the proposed formulation can be seen as a discontinuous dynamical
system (DDS) (Theorem \ref{thm:bt_dds}), with corresponding results regarding
existence and uniqueness (Theorem \ref{thm:filippov}). We provide sufficient
conditions under which a BT execution will converge to a desired region of the
state space (Theorem \ref{theorem:convergence}).

The organization of this letter is as follows.
In section \ref{sec:related-work}, we  discuss how our contributions differ from those presented in related work.
In section \ref{sec:preliminaries}, we provide a brief overview of tools for analyzing ordered trees and results regarding DDSs.
Then, in Section \ref{sec:bt-dynamical-system}, we formulate continuous-time BTs and connect them to DDSs in Section~\ref{sec:BT_as_DDS}.
Finally, in Section \ref{sec:convergence}, we present a convergence proof
and in Section \ref{sec:conclusion}, we state our conclusions.

\section{Related Work}\label{sec:related-work}
In this section, we will describe related work from a number of different aspects.

\noindent
\textbf{Continuous-time}: \enskip
In \cite{marzinotto2014towards},
a continuous-time BT is
informally described as a 
discrete-time BT with an infinite tick rate,
as a means to compare BTs to HDSs.
In \cite{klockner2014modelica}, instead of querying behaviors
at a certain tick rate,
behaviors run continuously and notify superior behaviors
when their status changes.
\emph{
Our work addresses the same problems;
however, our work does so on the basis
of a \textit{formal} state space definition of continuous-time BTs (Definition \ref{def:behavior}).
}

\noindent
\textbf{Hybrid dynamical Systems}: \enskip
The first comparison of BTs to HDSs appears to have
been made in \cite{ogren2012increasing}.
Therein, it was described how 
BTs  modularly represent HDSs
and implicitly encode 
explicit state transitions through its tree structure.
This discussion continued along the same lines in \cite{colledanchise2016behavior} and equivalence notions between discrete-time BTs and HDSs were presented in  \cite{marzinotto2014towards}.

In these works, the interpretation of an HDS is such that
a discrete state determines which
behavior to use.
However, as we will show, a BT is aptly
described by a DDS \cite{cortes2008discontinuous},
where the state's presence in certain regions 
solely determines which behavior is used.
\emph{
Thus, we go beyond related work by not only showing that BTs
more closely correspond to DDSs \cite{cortes2008discontinuous},
but we also do this \textit{formally} (Theorem \ref{thm:bt_dds}).
As a result, we also address  existence 
and uniqueness of solutions
(Theorem \ref{thm:filippov}).
}

\noindent
\textbf{Convergence analysis}: \enskip
It was shown in \cite{colledanchise2016behavior}
that the composition of behaviors in Fallback
BTs is similar to the idea
of sequential composition \cite{burridge1999sequential}.
Therein, sufficient conditions for
convergence to a goal state
were presented formally in terms of the attraction region of individual behaviors.
These concepts were applied in \cite{sprague2018adding} to guarantee 
BT performance in the presence of black-box controllers.

A version of BTs
called Robust Logical-Dynamical Systems was proposed in \cite{paxton2019representing},
which uses an Implicit-Sequence BT structure like in \cite{colledanchise2016behavior}.
Therein, they show convergence in the presence of uncontrolled behavior changes.
\emph{
Our work is related to all of the above in that we prove convergence in BTs (Theorem \ref{theorem:convergence});
however, our work is different in the sense that the results can be applied to  \textit{general} BT structures, not just special classes.
}

A concept of \cite{burridge1999sequential}
not used in the above works is the ``prepares graph'', 
a directed graph of transitions induced by the composition of policies.
In \cite{burridge1999sequential}, this graph is
used to construct a totally ordered subgraph of policies
that lead to the goal state.
This construction was extended in \cite{conner2006integrated}
to allow for multiple controllers in the subgraph to overlap
in order to attain more flexibility
in the presence of disturbances,
thereby forming a partially ordered subgraph. 
\emph{
We will use this notion of a prepares graph as a 
tool to prove the convergence of general BTs.
}

\section{Preliminaries}\label{sec:preliminaries}

In this section we will first describe how two partial orders can be used for analyzing ordered trees, and then present some results on DDSs.

\subsection{Ordered Trees}

As we will see below, BTs are ordered trees, and as was discussed in \cite{kuboyama2007matching}, ordered trees can either be seen as 
graphs,
as drawn in Fig.~\ref{fig:heater_bt}, or as a set of vertices with two partial orders, the so-called parent and sibling orders.

A directed graph is often defined in terms of $G=(V,E)$, where $V$ is the vertices and $E\subset V^2$ is the edges. If the graph has no cycles and no two distinct paths from a starting vertex meet at the same ending vertex, it is called a tree; if one vertex is designated as the root, it is called rooted. Given a root, the usual concepts of parent/child can be applied to each edge, with the parent being closer to the root and the child further away. To create an ordering between siblings (children of the same parent) the vertices can be embedded in a plane (as drawn on a paper) and the order given by clockwise or left/right positions.
In Fig.~\ref{fig:heater_bt}, the root would be vertex 0, and its two children vertex 1 and 4 (in that order) and so on.

In this letter, we will use the graph model
for BTs, but we will also make use of order theory for analyzing ordered trees, as described in \cite{kuboyama2007matching}.
As we will show, this formulation will support the analysis. 
We now use $(V,\leq_S, \leq_P)$ to define the tree, where $V$ is the vertex set as above, and $\leq_S, \leq_P$ are two partial orders on $V$, called the sibling and parent orders, respectively.

A partial order $\leq$ on a set is a homogeneous binary relation $\leq \subset V^2$ (if $(x,y)\in \leq$ we write $x \leq y$) that is 
reflexive ($\forall x \in V : x \leq x$), 
antisymmetric ($\forall x,y \in V : (x \leq y) \land (y \leq x) \implies x = y$), and
transitive ($\forall x,y,z \in V: (x \leq y) \land (y \leq z) \implies x \leq z$).
The order is partial, since two elements $x,y$ might not satisfy $x \leq y$ or $y \leq x$. If so, $x,y$ are said to be incomparable by $\leq$.
If all elements are comparable, the order is said to be a total order, instead of a partial order.
We write $x < y$ if $x \leq y$ and $x \neq y$, and for the reversed order $\geq$ we write $y \geq x$ if $x \leq y$.

In Fig.~\ref{fig:heater_bt}, we have that $1 \leq_S 4$, since $1$ and $4$ are siblings and $1$ is to the left of $4$. Note that $0$ and $1$ are incomparable by $\leq_S$, since they have no sibling relation. Instead, they are comparable by $\leq_P$, with $0 \leq_P 1$. Furthermore, 
$0$ and $3$ are comparable by $\leq_P$, with $0 \leq_P 3$ by transitivity, but
$0$ and $3$ are incomparable by $\leq_S$.

We can also combine orders into new orders as
\begin{equation}
    \leq_A \circ \leq_B := \left\{\left(x,z\right)\in V^2 \mid \exists y \in V: 
    \left(x \leq_A y\right) \land \left(y \leq_B z\right)\right\}.
\end{equation}
In this way, we can define a generalized uncle relation from the sibling and parent relations as 
$<_{LU} := <_S \circ \leq_P$ (left uncle)
$>_{RU} := >_S \circ \leq_P$ (right uncle). These relations include several  steps in both sibling and parent directions, thus including \blue{siblings, uncles}, great uncles, great-great uncles, and so on.
In Fig. \ref{fig:heater_bt}, we have that $4 >_{RU} 2$ and $4 >_{RU} 3$ because $4$ is a right uncle of $2$ and $3$.
    
Independently of the graph or ordered set representations, we will use the parent map $p : V \to V$, mapping a vertex to its parent.

\subsection{Dynamical systems theory}

In this section, we will remind readers of \blue{a result from} \cite{cortes2008discontinuous} on the existence and uniqueness of the solutions
to DDSs.
The notation used here will be used in the following sections
to show how BTs fit into this formalism.

\begin{theorem}[Existence and uniqueness {\cite[Proposition 5, p.53]{cortes2008discontinuous}}]
        \label{theorem:cortes_p5}
        Let $X:\mathbb{R}^n \rightarrow \mathbb{R}^n$ be a piecewise continuous vector field, 
        with $\mathbb{R}^n = D_1 \cup D_2$.
        Let $S_X=\partial D_1 = \partial D_2$, where $\partial$ is the boundary operator, 
        be the set of points at which X is discontinuous, 
        and assume that $S_X$ is a $C^2$-manifold.
        Furthermore, assume that, for $i \in \{1,2\}$,
        $X_{|\bar D_i}$ is continuously differentiable on $D_i$ and
        $X_{|\bar D_1}-X_{|\bar D_2}$ is continuously differentiable on $S_X$,
        where $X_{|\bar D_i}$ is the continuous extension of the restriction of $X$ to $\bar D_i$.
        If, for each $x\in S_X$,
        either $X_{|\bar D_1}$ points into $D_2$ 
        or $X_{|\bar D_2}$ points into $D_1$,
        there will exist a unique Filippov 
        solution to $\dot x = X(x)$ starting from each initial condition.
    \end{theorem}

    \section{Continuous-time BTs}\label{sec:bt-dynamical-system}

    In this section, we will define continuous-time BTs, and see how the example of
    Fig.~\ref{fig:heater_bt} forms a continuous-time controller.
    
    As noted above, BTs are a hierarchical and modular way of combining controllers into new controllers.
    In this letter we let all controllers be state-feedback controllers, i.e. functions from the state space $\mathbb{R}^n$ to some control space $\mathbb{R}^m$. 
    If one wants to include some internal dynamics, such as a Kalman filter, in the controller, the state space can be extended.
    
    \begin{definition}[Behavior Tree]\label{def:behavior}
        A function 
         $\bt_i: \mathbb{R}^n \to \mathbb{R}^m \times \{\mathcal{R},\mathcal{S},\mathcal{F}\}$,
         defined as
        \begin{equation}\label{eq:behavior}
            \bt_i(x):=\left(u_i\left(x\right),r_i\left(x\right) \right),
        \end{equation}
        where $i \in V$ is an index,
        $u_i: \mathbb{R}^n \rightarrow  \mathbb{R}^m$ is a controller,
        and $r_i: \mathbb{R}^n \rightarrow  \{\mathcal{R},\mathcal{S},\mathcal{F}\}$ 
        is a metadata function, describing the progress of the controller in terms of the 
        outputs:
        running ($\mathcal{R}$),
        success ($\mathcal{S}$), 
        and
        failure ($\mathcal{F}$).
        Define the metadata regions for $x \in \mathbb{R}^n$ as the running, success, and failure regions:
        \begin{equation}\label{eq:metadata-regions}
    \begin{gathered}
        R_i :=\left\{x: r_i(x)=\mathcal{R} \right\}, \\
        S_i :=\left\{x: r_i(x)=\mathcal{S} \right\}, \quad F_i :=\left\{x: r_i(x)=\mathcal{F} \right\},
    \end{gathered}
        \end{equation}
        respectively,
        which are pairwise disjoint and cover $\mathbb{R}^n$.
    \end{definition}

    The metadata can intuitively be interpreted as follows. 
    If $x\in S_i$, $\bt_i$ has either succeeded with whatever it was supposed to do (such as opening a door), or the goal was already achieved to begin with (the door was open).
    Either way, it might make sense to execute another controller to achieve some other goal (perhaps a goal that was intended to be achieved after opening the door).
    
    If $x\in F_i$, $\bt_i$ has either failed (the door to be opened turned out to be locked), or has no chance of succeeding (the door is out of reach from the current position).
    Either way, it might make sense to execute another controller
    (either to open the door in some other way or to achieve a higher-level goal in a way that does not involve opening the door).
    
    If $x\in R_i$, it is too early to determine if $\bt_i$ will succeed or fail. In most cases, it makes sense to continue executing $\bt_i$, but it could also be reasonable to change the controller if some other action is more important (e.g. low battery level indicates the need for recharging).
    
    \begin{definition}[Continuous BT execution]\label{def:bt_execution}
        Given some dynamical system $f : \mathbb{R}^n \times \mathbb{R}^m \to \mathbb{R}^n$ that is to be controlled, and assuming the root of the BT is $\bt_0$ (has index 0), we have
        \begin{equation}
            \label{eq:bt_execution}
            \dot x = f\left(x,u_0\left(x\right)\right),
        \end{equation}
        where $u_0(x)$ is given by (\ref{eq:behavior}).
    \end{definition}
    Below we will describe the properties of this execution, and in particular show that it can be seen as a DDS, with corresponding results regarding the existence and uniqueness of solutions.

    As described above, knowing if a lower-level controller failed, succeeded, or is still trying (running) is crucial for a higher-level controller to decide 
    if another sequence should be initiated,
    or if some kind of fallback action needs to be invoked to achieve the desired outcome. 
    These two cases are captured by the two fundamental BT composition types: Sequence and Fallback.
    The result of these behavior compositions
    is simply another BT that satisfies (\ref{eq:behavior}).
    This is what gives BTs their hierarchical modularity.

    A Sequence is used to combine subtrees that are to be executed in order, 
    where each one requires the \emph{success} of the previous action. If any subtree fails, the whole sequence fails.
    In Fig.~\ref{fig:kitchen_lamp}, node $0$ is a Sequence. First, node 1 is executed to get into the kitchen, and then node $2$ is executed to turn one of the lamps on. But it only makes sense to try turning the lamps on if the action of moving to the kitchen succeeds.
    Formally, a Sequence is defined as follows.

    \begin{wrapfigure}[9]{r}{0.42\columnwidth}
        \centering
        \footnotesize
              \begin{forest}
                  for tree={
                      minimum height=2em,
                      minimum width=2em,
                      inner sep=0.3pt,
                  }
                  [$\rightarrow$, draw, tikz={\node[left=1pt of .west]  {$0$};},
                      [{$
                          \begin{array}{l}
                              \text{Go to}\\
                              \text{Kitchen}
                          \end{array}
                          $}, draw, tikz={\node[left=1pt of .west] {$1$};}],
                      [$?$, draw, tikz={\node[above=1pt of .north]  {$2$};},
                          [{$
                          \begin{array}{l}
                              \text{Turn on}\\
                              \text{Lamp A}
                          \end{array}
                          $}, draw, tikz={\node[left=1pt of .west]  {$3$};}],
                          [{$
                          \begin{array}{l}
                              \text{Turn on}\\
                              \text{Lamp B}
                          \end{array}
                          $}, draw, tikz={\node[above=1pt of .north]  {$4$};}]
                      ]
                  ]
              \end{forest}
              \caption{}
              \label{fig:kitchen_lamp}
      \end{wrapfigure}

      \begin{definition}[Sequence]
        A function
        $Seq$ 
        that composes an arbitrarily finite sequence of 
        $M \in \mathbb{N}$  BTs
        into a new BT as
        \begin{equation}\label{eq:sequence}
            Seq\left[\bt_1, \dots, \bt_M\right]\left(x\right) :=
            \begin{cases}
            \bt_M\left(x\right) & \text{if} \quad x \in S_1 \cap \dots S_{M-1} \\
            \vdots & \vdots \\
            \bt_2\left(x\right) & \text{else-if} \quad  x \in S_1 \\
            \bt_1\left(x\right) & \text{else}.
            \end{cases}
        \end{equation}
        If $\bt_i = Seq[\bt_1, \dots, \bt_M]$,
        then $j,k \in \{1, \dots, M\}$
        are the children of $i$,
        such that $p(j) = i$,
        and are related as siblings, by $j \leq_S k$, if $j \leq k$.
    \end{definition}
    As can be seen in (\ref{eq:sequence}), a subtree $\bt_i$ is only executed if the state is in the success region of the siblings to the left $\bt_j, j<i$.
    
    A Fallback on the other hand only executes the next subtree if the previous one \emph{fails}. If any subtree succeeds, the Fallback returns success, but it only returns failure if all subtrees fail.
    In Fig.~\ref{fig:kitchen_lamp}, node \blue{$2$} is a Fallback, and the two subtrees correspond to turning on either lamp A or lamp B.
    
    \begin{definition}[Fallback]
        A function
        $Fal$ 
        that composes an arbitrarily finite sequence of 
        $M \in \mathbb{N}$  BTs
        into a new BT as
        \begin{equation}\label{eq:fallback}
            Fal\left[\bt_1, \dots, \bt_M\right]\left(x\right) :=
            \begin{cases}
            \bt_M\left(x\right) & \text{if} \quad x \in F_1 \cap \dots F_{M-1} \\
            \vdots & \vdots \\
            \bt_2\left(x\right) & \text{else-if} \quad  x \in F_1 \\
            \bt_1\left(x\right) & \text{else}.
            \end{cases}
        \end{equation}
        If $\bt_i = \blue{Fal}[\bt_1, \dots, \bt_M]$,
        then $j,k \in \{1, \dots, M\}$
        are the children of $i$,
        such that $p(j) = i$,
        and are related as siblings, by $j \leq_S k$ if $j \leq k$.
    \end{definition}
    
    The metadata regions (\ref{eq:metadata-regions}) of the Sequence and Fallback compositions are given by the definition, but can also be explicitly computed in terms of the children regions and the orders $<_S, <_P$ as follows.
    
    \begin{lemma}\label{lemma:metadata-sequence}
        \blue{The metadata regions of a Sequence $\bt_i$ can be computed from the children metadata regions as follows:}
        
        \begin{equation}\label{eq:sequence-metadata}
        \begin{gathered}
            R_i =\bigcup_{p\left(j\right) = i}\left( R_j \bigcap_{k <_S j} S_k\right), \\
            S_i = \bigcap_{p\left(j\right) = i} S_j,\quad F_i = \bigcup_{p\left(j\right) = i}\left( F_j \bigcap_{k <_S j} S_k \right).
        \end{gathered}
        \end{equation}
    \end{lemma}
    \begin{proof}
        A straightforward application of 
        (\ref{eq:metadata-regions})
        and (\ref{eq:sequence}). The running region of the sequence is the running region of the first child and the intersection of the success region of the first child with the running region of the second child and so on. The failure region works similarly, whereas the success region is the intersection of all the children success regions, as the sequence requires all children to succeed to return success.
    \end{proof}
    
    \begin{lemma}\label{lemma:metadata-fallback}
        \blue{The metadata regions of a Fallback $\bt_i$ can be computed from the children metadata regions as follows}
            \begin{equation}\label{eq:fallback-metadata}
            \begin{gathered}
            R_i = \bigcup_{p\left(j\right) = i}\left( R_j \bigcap_{k <_S j} F_k\right), \\
            S_i = \bigcup_{p\left(j\right) = i}\left( S_j \bigcap_{k <_S j} F_k \right), \quad
            F_i = \bigcap_{p\left(j\right) = i} F_j.
            \end{gathered}
        \end{equation}
    \end{lemma}
    \begin{proof}
        A straightforward application of 
        (\ref{eq:metadata-regions})
        and (\ref{eq:fallback}). 
        The running region is similar as for the Sequence above. The success region is similar to the running region, but the failure region is different since it requires all children to fail before returning failure.
    \end{proof}
    
    \section{BTs as discontinuous dynamical systems}
    \label{some_label}
    \label{sec:BT_as_DDS}
    
    We need to show that the BT execution of (\ref{eq:bt_execution}) can be seen as a DDS.
    Thus we need to identify the operating regions $\Omega_i$ of the BT, i.e. the regions where the root BT executes a particular subtree $\bt_0=\bt_i$.
    As we will see, the $\Omega_i$ will depend on both the subtree $\bt_i$ itself, and its place in the surrounding BT.
    But, before we can define the operating region $\Omega_i$ we need to define the influence region $I_i$ and the success and failure pathways $\mathfrak{S}, \mathfrak{F}$.
    
    Informally, the influence region $I_i$ is the region where the design of $\bt_i$ influences the execution of $\bt_0$, either by returning e.g. failure so another node executes or by executing itself (thus we will have $I_i \supset \Omega_i$).
    
    We will be using the so-called left uncle (LU) order
    $<_{LU} := <_S \circ \leq_P$ defined in Section \ref{sec:preliminaries}. Note that $\bt_j : j<_{LU} i$ are left siblings of either $i$ or any ancestors of $i$. For a state to be in $I_i$ it needs to be in the success region of the left uncles that have a Sequence as a parent, and in the failure region of the left uncles that have a Fallback as a parent.
    Formally we write the following.

    \begin{definition}[Influence Region] \label{def:influence}
        A subset of the state space defined for $\bt_i$ as
        \begin{equation}\label{eq:influence}
            I_i := 
            \underset{\substack{j <_{LU} i \\ \bt_{p\left(j\right)} is \mbox{ } Seq }\quad}{\bigcap S_j}
            \cap
            \underset{\substack{j <_{LU} i \\ \bt_{p\left(j\right)} is \mbox{ } Fal.}\quad}{\bigcap F_j}
        \end{equation}
    \end{definition}
    
    In the example of Fig.~\ref{fig:kitchen_lamp}, assuming the state space is $\mathbb{R}^n$, 
    we have that $I_0=\mathbb{R}^n$, $I_1=\mathbb{R}^n$,  $I_2=S_1$, $I_3=S_1$, and $I_4=S_1 \cap F_3$. 
    Thus, a change in $\bt_1$ can influence  $\bt_0$ in any part of the state space, but a change in $\bt_4$ can only influence $\bt_0$ if $x\in S_1 \cap F_3$, 
    i.e., if going to the kitchen was successful and turning on lamp A failed.
    
    If the state is in $I_i$ and $\bt_i$ returns running, it will execute. 
    But, it will also execute in the case when $\bt_i$ returns success or failure and that same metadata is progressed all the way up to the root. Thus we need to identify what subtrees are on the so-called success and failure pathways. We now make use of the right uncle (RU) order that was also defined in Section~\ref{sec:preliminaries}, $>_{RU} := >_S \circ \leq_P$. Similarly, $\bt_j : j>_{RU} i$ are right siblings of either $i$ or any ancestors of $i$.
    
    Informally, success pathways are vertices $i$ such that there are no right uncles, with Sequence parents, that can take over the execution when $\bt_i$ returns success.
    Similarly, failure pathways are  vertices $i$ such that there are no right uncles, with Fallback parents, that can take over the execution when $\bt_i$ returns failure.
    We call them pathways since if $i$ is on the pathway then so is every other vertex on the path from $i$ to the root.
    Formally, we write the following.
    
    \begin{definition}[Success and failure pathways]
        \begin{align}
            \label{eq:success-pathway}
            \mathfrak{S} &:= \left\{i \in V \mid 
            \not \exists j \in V : \left(j >_{RU} i\right) \land 
            \left(\bt_{p(j)} ~ is ~ Seq\right)\right\} \\
            \label{eq:failure-pathway}
            \mathfrak{F} &:= \left\{i \in V \mid 
            \not \exists j \in V: \left(j >_{RU} i\right) \land 
            \left(\bt_{p(j)} ~ is ~ Fal\right)\right\},
        \end{align}
        respectively.
    \end{definition}
    
    In the example of Fig.~\ref{fig:kitchen_lamp}, we have that 
    $\mathfrak{S}=\{0,2,3,4\}$, since success from these nodes leads to success of the entire BT, and only success in going to the kitchen leads to other actions.
    Similarly,
    $\mathfrak{F}=\{0,1,2,4\}$, since failure from these nodes leads to failure of the entire BT, and only a failure in turning on lamp A can be handled (by turning on lamp B).
    
    We are now ready to define the operating regions.
    \begin{definition}[Operating Region]
        A subset of the state space
        defined for $\bt_i$ as
        \begin{equation}\label{eq:operating}
        \Omega_i := 
        \begin{cases}
            I_i \cap (R_i \cup S_i \cup F_i) = I_i & \text{if} \quad i \in \mathfrak{S} \cap \mathfrak{F} \\
            I_i \cap (R_i \cup S_i) & \text{else-if} \quad i \in \mathfrak{S} \\
            I_i \cap (R_i \cup F_i) & \text{else-if} \quad i \in \mathfrak{F} \\
            I_i \cap R_i  & \text{else}.
        \end{cases}
        \end{equation}
    \end{definition}
    
    In the example of Fig.~\ref{fig:kitchen_lamp}, we have that 
    $\Omega_0=\mathbb{R}^n$, 
    $\Omega_1=R_1 \cup F_1$, 
    $\Omega_2=S_1 \cap (R_2 \cup S_2)$,
    $\Omega_3=S_1$, 
    $\Omega_4=S_1 \cap F_3$. 
    
    
    
    We will now show that a BT's operating region is partitioned by its childrens' operating regions.
    
    \begin{lemma}\label{lemma:operating-partition}
        Operating regions 
        of siblings are 
        pairwise disjoint,
        $\Omega_i \cap \Omega_j = \emptyset$ for all $i <_S j$,
        and cover their parent's operating region,
        $\Omega_i = \bigcup_{p(j) = i} \Omega_j$.
    \end{lemma}
    \begin{proof}
    As shown in \cite{colledanchise2016behavior}, 
    compositions can be expressed as follows:
    $Seq[\bt_1, \bt_2, \bt_3] = Seq[\bt_1, Seq[\bt_2, \bt_3]]$
    and $Fal[\bt_1, \bt_2, \bt_3] = Fal[\bt_1, Fal[\bt_2, \bt_3]]$.
    Thus, it is sufficient to analyze the case of two children.
    
    Let $0, 1, 2 \in V$ such that $1 <_S 2$ and $p(1) = p(2) = 0$.
    We will now apply
    each case of (\ref{eq:operating}) to $\Omega_1$,
    assuming $2 \in \mathfrak{S} \cap \mathfrak{F}$,
    which implies $0 \in \mathfrak{S} \cap \mathfrak{F}$ according
    to (\ref{eq:success-pathway})
    and (\ref{eq:failure-pathway}). 
    
    The first case is ruled out because 
    $1 \in \mathfrak{S} \cap \mathfrak{F}$ implies that
    $\not \exists j: j >_{RU} 1$ 
    and we know that $2 >_{RU} 1$.
    
    
    In the second case, 
    $1 \in \mathfrak{S} \setminus \mathfrak{F}$ implies that $\not \exists j : (j >_{RU} 1) \land (\bt_{p(j)}~is~Seq)$,
    thus node $0$ must be a Fallback.
    With the application of (\ref{eq:fallback-metadata}), (\ref{eq:influence}), and (\ref{eq:operating}),
    we then have 
    $\Omega_0 = I_0$, 
    $\Omega_1 = I_1 \cap (R_1 \cup S_1) = I_0 \cap (R_1 \cup S_1)$, 
    and
    $\Omega_2 = I_2 = I_0 \cap F_1$.
    From this, we see that
    $\Omega_1 \cap \Omega_2 = I_0 \cap (R_1 \cup S_1) \cap I_0 \cap F_1 = \emptyset$
    because $\{R_1, S_1, F_1\}$ are pairwise disjoint by (\ref{eq:metadata-regions}).
    Additionally, 
    $\Omega_1 \cup \Omega_2 = (I_0 \cap (R_1 \cup S_1 ))
    \cup (I_0 \cap F_1)
    = 
    I_0 \cap(R_1 \cup S_1 \cup F_1)
    = I_0
    = \Omega_0$.

    In the third case, 
    $1 \in \mathfrak{F} \setminus \mathfrak{S}$ implies that  $\not \exists j : (j >_{RU} 1) \land (\bt_{p(j)}~is~Fal)$ 
    thus node $0$ must be a Sequence.
    With the application of (\ref{eq:sequence-metadata}), (\ref{eq:influence}), and (\ref{eq:operating}),
    we then have 
    $\Omega_0 = I_0$, 
    $\Omega_1 = I_1 \cap (R_1 \cup F_1) = I_0 \cap (R_1 \cup F_1)$, 
    and
    $\Omega_2 = I_2 = I_0 \cap S_1$.
    From this, we see that
    $\Omega_1 \cap \Omega_2 = I_0 \cap (R_1 \cup F_1) \cap I_0 \cap S_1 = \emptyset$
    because $\{R_1, S_1, F_1\}$ are pairwise disjoint by (\ref{eq:metadata-regions}).
    Additionally, 
    $\Omega_1 \cup \Omega_2 = (I_0 \cap (R_1 \cup F_1 ))
    \cup (I_0 \cap S_1)
    = 
    I_0 \cap(R_1 \cup S_1 \cup F_1)
    = I_0
    = \Omega_0$.

    The fourth case's proof follows similarly with $\Omega_1 = I_1 \cap R_1$.
    The proofs for the cases of (\ref{eq:operating}) for $\Omega_2$ are also similar.
    \end{proof}
    
    We will now formally prove that the state's presence in $\Omega_i$ is indeed a \blue{sufficient} condition to
    conclude that $\bt_i$ is being executed.
    \begin{theorem}\label{thm:bt_dds}
        Let $P$ be the set of leaf nodes whose operating regions are non-empty:
        \begin{equation} \label{eq:subsystem-nodes}
            P := \left\{ i \in V \mid 
        \left(\Omega_i \neq \emptyset\right) \land 
        \left(\not \exists j \in V : j >_P i\right) \right\}.
        \end{equation}
        Then, we have
        $x \in \Omega_i: i \in P \implies \dot{x} = f(x,u_0(x))=f(x,u_i(x))$
        and $\underset{i \in P }{\bigcup }\Omega_i = \mathbb{R}^n$.
    \end{theorem}
    \begin{proof}
        We need to show that
        $x \in \Omega_i: i \in P \implies f(x,u_0(x))=f(x,u_i(x))$ 
        and that
        $\{\Omega_i\}_{i \in P}$ 
        cover the state space.
    
        We have that $\Omega_i \subset I_i$ by (\ref{eq:operating}) and from (\ref{eq:influence}) we see that no leaf to the left of $u_i$ can execute. Furthermore, by the construction of (\ref{eq:operating}), either $x \in R_i$, or $x$ is in the success or failure region of a node on a success or failure pathway (respectively), so no leaf to the right of $u_i$ can execute. 
        Thus, we conclude that
        $x \in \Omega_i: i \in P \implies f(x,u_0(x))=f(x,u_i(x))$.
    
        Now we need to show that 
        $\{\Omega_i\}_{i \in P}$ 
        cover the state space.
        From Lemma \ref{lemma:operating-partition}
        we have that $\Omega_i$ for a set of siblings are pairwise disjoint and cover $\Omega_{p(i)}$. 
        By definition, $I_0 = \mathbb{R}^n$ and since $0 \in \mathfrak{S} \cap \mathfrak{F}$ we have 
        $\Omega_0=I_0=\mathbb{R}^n$ by (\ref{eq:operating}). Applying Lemma 
        \ref{lemma:operating-partition} recursively down the tree we see that for the leaves in $P$ we have that
        $\{\Omega_i\}_{i \in P}$ are pairwise disjoint and  cover $\mathbb{R}^n$, $\underset{i \in P }{\bigcup }\Omega_i = \mathbb{R}^n$.
    \end{proof}
    
    \begin{theorem}\label{thm:filippov}
        The execution (\ref{eq:bt_execution}) will have a unique Filippov solution \blue{(see \cite{cortes2008discontinuous})} for each initial state
        if, for every pair of neighboring sets with index in $P$, i.e. sets $\Omega_i,\Omega_j$ with $i,j\in P$ and $\partial \Omega_i \cup \partial \Omega_j\neq \emptyset$, the sets $\Omega_i,\Omega_j$ and the vector field
        \begin{equation}
        X(x) = 
            \begin{cases}
            f\left(x,u_i(x)\right) & \text{if}  \quad x\in \Omega_i \\
            f\left(x,u_j(x)\right) &\text{else}
        \end{cases}
        \end{equation}
        are such that the following holds with $D_1=\Omega_i$ and $D_2=\mathbb{R}^n \setminus D_1$.
        $S_X= \partial D_i$ is the set where $X(x)$ is discontinuous
        and $S_X$ is a $C^2$-manifold.
        Furthermore, for $i \in \{1,2\}$,
        $X_{|\bar D_i}$ is continuously differentiable on $D_i$ and
        $X_{|\bar D_1}-X_{|\bar D_2}$ is continuously differentiable on $S_X$.
        For each $x\in S_X$,
        either $X_{|\bar D_1}$ points into $D_2$ 
        or $X_{|\bar D_2}$ points into $D_1$.
    \end{theorem}
    \begin{proof} 
    A straightforward application of
    Theorem~\ref{theorem:cortes_p5} for every neighboring pair of $\Omega_i$.
    \end{proof}

        Sufficient conditions for the existence and uniqueness of BT executions can thus be found using the corresponding results for DDS in Theorem~\ref{theorem:cortes_p5}.
    
\pagebreak
    \section{Convergence analysis} \label{sec:convergence}
\begin{wrapfigure}[11]{r}{0.29\columnwidth}
  \centering
  \includegraphics[width=0.29\columnwidth]{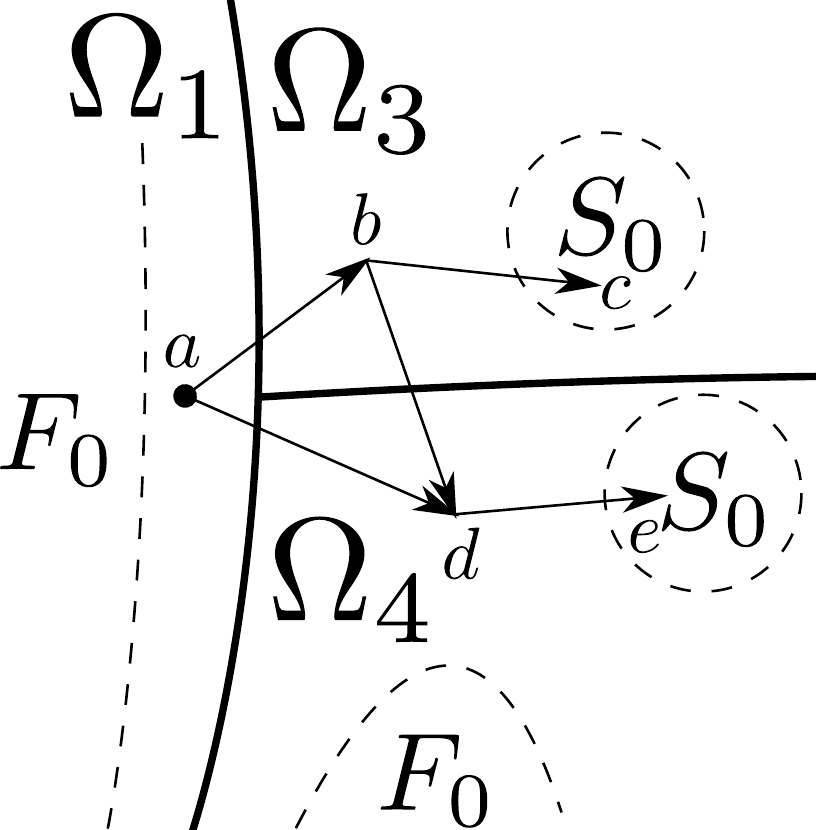}
    \caption{Prepares graph for the BT in Fig. \ref{fig:kitchen_lamp}.}
    \label{fig:prepares}
\end{wrapfigure}
In this section, we will state the conditions under which
a general BT 
is convergent.
The main idea of our convergence theorem is similar  
to the concept of \emph{prepares} from \cite{burridge1999sequential}.
Given a BT and its operating regions, the region of attraction of each policy invokes switching between operating regions,
thereby inducing a partial order $\leq_f$ of transitions.

The reflexive-transitive reduction of this partial order is a directed acyclical graph (prepares graph),
as illustrated in Fig. \ref{fig:prepares} for the kitchen-lamp example in Section \ref{sec:bt-dynamical-system}.
The transitions (edges) of this graph are described as follows: 
$(a, b)$ going to the kitchen and trying to turn on lamp A because it is closer,
$(a, d)$ going to the kitchen and trying to turn on lamp B because it is closer,
$(b, d)$ trying to turn on lamp B because lamp A did not work,
$(b, c)$ successfully turning on lamp A,
$(d, e)$ successfully turning on lamp B.
Note, the dashed regions in Fig. \ref{fig:prepares} correspond to the success and failure pathways. 
Informally speaking, the BT will be convergent if this graph is acyclical and has all its sinks in success regions.
We will now formally state the convergence theorem.

\begin{theorem}\label{theorem:convergence}
    If there exists a subset $L \subseteq P$ and a partial order $\leq_f \subset L^2$
    such that the constraint region
    \begin{equation}\label{eq:constraint}
        \Lambda_i := \bigcup_{j \geq_f i} \Omega_j \setminus F_0
    \end{equation}
    is invariant under $f(x, u_i(x))$ for all $i \in L$,
    and there exists a finite time $\tau_i > 0$,
    such that if 
    $x(t) \in \Omega_i \setminus S_0$
    then 
    $x(t + \tau_i) \not \in \Omega_i \setminus S_0$
    for all $i \in L$,
    then there exists a maximum number of transitions
    $N \in \mathbb{N}$ and a maximum duration $t' > 0$,
    such that if $x(0) \in \Lambda_i$ for any $i \in L$,
    then 
    $x(t) \in S_0$ 
    in bounded time $t \leq t'$ within $N$ transitions.
\end{theorem}
\begin{proof}
    We have that if $x(t) \in \Omega_i \setminus S_0$ then $x(t + \tau_i) \not \in \Omega_i \setminus S_0$.
    But, $\Lambda_i$ is invariant under $f(x, u_i(x))$.
    Thus, if $x(t) \in \Lambda_i$ then $x(t + \tau_i) \in \Omega_j \setminus F_0$ for some  $j \geq_f i$, meaning that either $x(t + \tau_i) \in R_0$ or $x(t + \tau_i) \in  S_0$.
    Thus, if $x(0) \in \Lambda_i$ then $x(t) \in S_0$ in bounded time $t \leq t'$ with $t' = \max_{L_0 \subseteq L} \sum_{k \in L_0} \tau_k$ and at most $N = \max_{L_1 \subseteq L} \mid L_1 \mid$ transitions, such that $L_0, L_1$ are maximal, totally ordered by $\leq_f$, and $i \leq_f k$ for all $k \in L_0 \cup L_1$.
    In other words, $L_0$ and $L_1$ are the chains of transitions with the largest duration and cardinality, respectively.
\end{proof}

We now have a tool to assess the convergence properties of a general BT.
The key challenge  is thus to design the structure of the BT itself and its controllers to satisfy Theorem \ref{theorem:convergence}.
An extended version of this paper, with a longer example of the application of this result can be found in 
\cite{sprague2021continuous}.

\section{Example}

In this section we will illustrate Theorem \ref{theorem:convergence} with a simple example.

     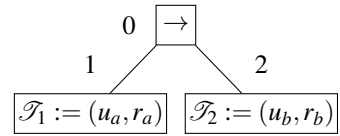
\begin{figure}[t]
            \centering
            \begin{forest}
                for tree={
                    minimum size=1.5em,
                    inner sep=1pt,
                    l=1.2cm
                }
                [$\rightarrow$, draw, tikz={\node[left=4pt of .west]  {$0$};}
                    [{$\bt_1 := \left(u_a, r_a \right)$}, draw, tikz={\node[above=4pt of .north]  {$1$};}]
                    [{$\bt_2 := \left(u_b, r_b \right)$}, draw, tikz={\node[above=4pt of .north]  {$2$};}]
                ]
            \end{forest}
            \caption{
                A BT controller (\ref{eq:pendulum-bt}) for the inverted pendulum (\ref{eq:pendulum}).
            }
            \label{fig:pendulum-bt}
        \end{figure}
        
    \begin{figure}[t]
        \centering
        \includegraphics[width=\columnwidth]{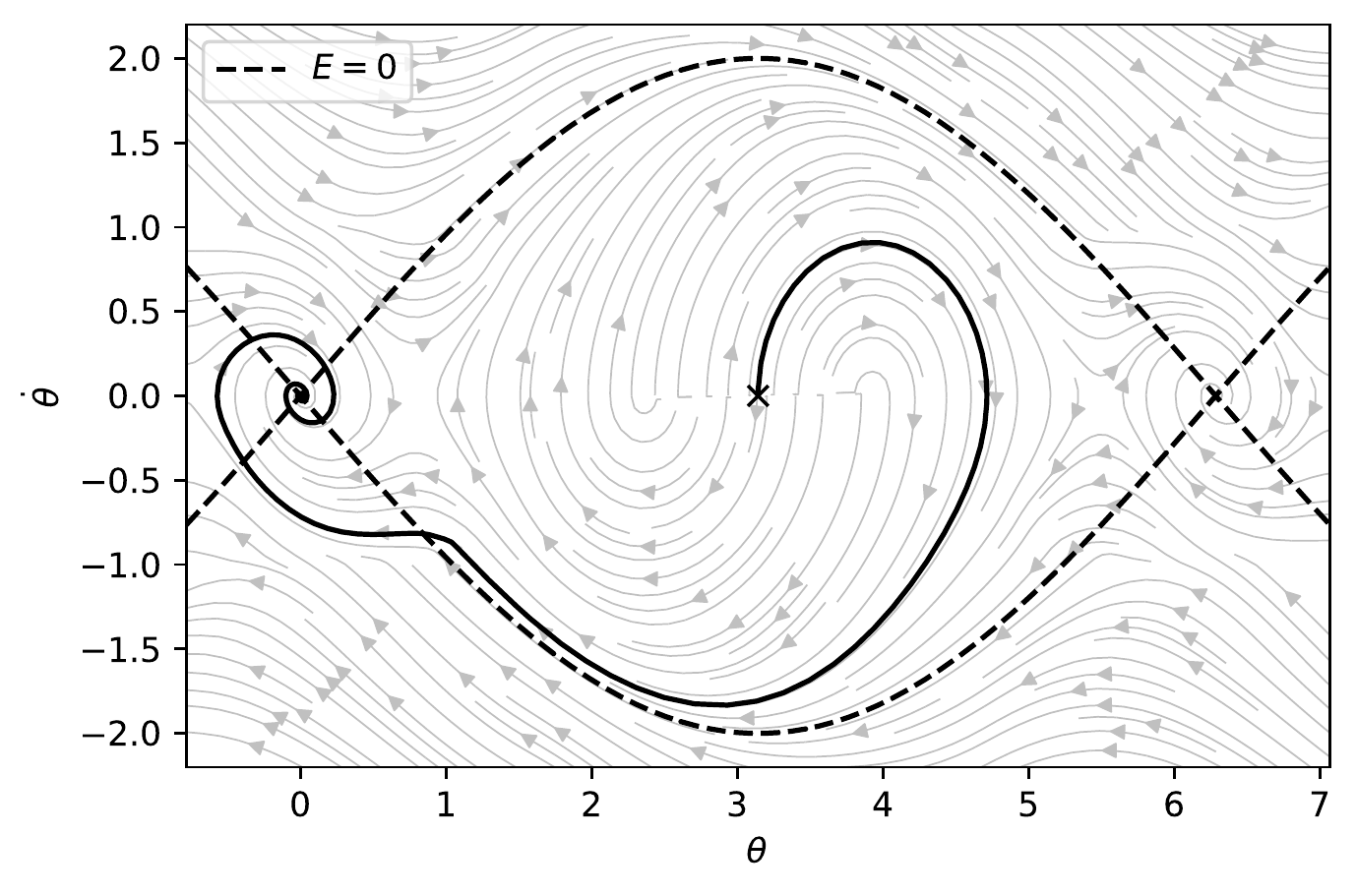}
        \caption{The piecewise-continuous vector field formed by the BT in Fig.  \ref{fig:pendulum-bt}.}
        \label{fig:pendulum-field}
    \end{figure}

Consider the normalized inverted pendulum model from \cite{aastrom2000swinging}
        \begin{equation} \label{eq:pendulum}
            \dot{x} = f\left(x, u\right) = 
            \left[\begin{matrix}\dot{\theta}\\\ddot{\theta}\end{matrix}\right] =
            \left[\begin{matrix}\dot{\theta}\\ \sin{\left(\theta \right)} - u \cos{\left(\theta \right)} \end{matrix}\right]
        \end{equation}
        where $\theta \in \mathbb{R}$ is the pendulum's angle from the vertical,
        $x := [\theta, \dot{\theta}]^\intercal$ is the state,
        and $u \in \mathbb{R}$ is a control input.
        We want to stabilize (\ref{eq:pendulum}) to the unstable equilibrium at stationary-upright configuration, where $\cos(\theta) = 1$ and $\dot{\theta} = 0$.

        A popular technique for doing so is energy control.
        Following \cite{aastrom2000swinging}, define the energy of (\ref{eq:pendulum}) as
        \begin{equation}
            E := \frac{\dot{\theta}^{2}}{2} + \cos{\left(\theta \right)} - 1
        \end{equation}
        and a control policy with constants $k_E, u_m \in \mathbb{R}_{>0}$ as
        \begin{equation} \label{eq:uE}
            u_a\left(x\right) := \text{sat}_{u_m} \left( k_E E \text{sgn}\left(\dot{\theta} \cos\left(\theta\right)\right)\right),
        \end{equation}
        where $\text{sat}_{u_m}$ ensures that $u_a(x) \in [-u_m, u_m]$.
        Policies such as (\ref{eq:uE}) are well-known to exponentially stabilize the
        pendulum (\ref{eq:pendulum}) to its homoclinic orbit about $E = 0$ (shown by the dashed lines in Fig. \ref{fig:pendulum-field})
        starting from all states other than the stable stationary-downward configuration,
        where $\cos(\theta) = -1$ and $\dot{\theta} = 0$. 

        Unfortunately, however, the stationary-upright configuration is only a saddle equilibrium of (\ref{eq:pendulum})
        under the influence of (\ref{eq:uE}).
        Thus, the system would only periodically pass through the stationary-upright configuration.
        Therefore, we need to define a local controller to ``switch on'' and stabilize the system when close enough to the stationary-upright configuration.
        Define a linear-feedback policy with constants $k_{\theta}, k_{\dot{\theta}} \in \mathbb{R}_{>0}$ as
        \begin{equation}\label{eq:linear-control}
            u_b\left(x\right) := k_\theta\sin\left(\theta\right) + k_{\dot{\theta}} \dot{\theta}.
        \end{equation}
        The policy (\ref{eq:linear-control}) is exponentially stabilizing within some region
        of the statespace around $\sin(\theta) = 0$ and $\dot{\theta} = 0$, which is met at both the stationary-upright and stationary-downward configurations.

        To make sure that the controller is only used in the stationary-upright configuration, we define an error metric
        \begin{equation}
            \delta\left(x\right) := \sqrt{l_\theta \left(\cos\left(\theta\right) - 1\right)^2 + l_{\dot{\theta}}\dot{\theta}^2},
        \end{equation}
        with constants $l_\theta, l_{\dot{\theta}} \in \mathbb{R}_{>0}$,
        which will only be zero at the stationary-upright configuration.
        Using this metric for the policies above, we define their metadata functions and regions,
        \begin{equation}
            r_a\left(x\right) := 
            \begin{cases}
                \mathcal{S} & \text{if} ~ \delta\left(x\right) \leq \epsilon_a \\
                \mathcal{R} & \text{else}    
            \end{cases}
            \quad
            \begin{aligned}
                R_a &= \left\{x : \delta \left(x\right) > \epsilon_a\right\} \\
                S_a &= \left\{x : \delta \left(x\right) \leq \epsilon_a\right\} \\
                F_a &= \emptyset
            \end{aligned}
        \end{equation}
        and 
        \begin{equation}
            r_b\left(x\right) := 
            \begin{cases}
                \mathcal{S} & \text{if} ~ \delta\left(x\right) \leq \epsilon_b \\
                \mathcal{R} & \text{else}
            \end{cases}
            \quad
            \begin{aligned}
                R_b &= \left\{x : \delta \left(x\right) > \epsilon_b\right\} \\
                S_b &= \left\{x : \delta \left(x\right) \leq \epsilon_b\right\} \\
                F_b &= \emptyset
            \end{aligned},
        \end{equation}
        where $\epsilon_a, \epsilon_b \in \mathbb{R}_{>0}$ are constants such that $\epsilon_b \leq \epsilon_a$, and $S_a$ is a positively invariant set
        of $f(x, u_b(x))$ containing its equilibrium (the stationary-upright configuration).
        

        Since the energy-based policy $u_a$ is exponentially stabilizing to the zero-energy manifold,  we have that
        $\left| E\left(t\right) \right| \leq \left|E\left(0\right)\right| \text{e}^{-\alpha t}$
        for some constant $\alpha \in \mathbb{R}_{>0}$.
        This implies that, for all $\epsilon'_a \in \mathbb{R}_{>0}$,
        there exists $\tau'_a \in \mathbb{R}_{>0}$ such that
        $\left| E\left(\tau'_a\right)  \right| \leq \left|E\left(0\right)\right| \text{e}^{-\alpha \tau'_a} < \epsilon'_a$.
        Since the angular velocity $\dot{\theta}$ maintains positivity or negativity for the duration of each orbit (see Fig. \ref{fig:pendulum-field}),
        it is implied that, if $S_a \subseteq \{x : E(x) \leq \epsilon'_a\}$,
        then there exists a finite bound $\tau_a \in \mathbb{R}_{\geq \tau'_a}$,
        such that, if $x(0) \in R_a$ then $x(t) \in S_a$ in finite time $t \leq \tau_a$
        for the execution $\dot{x} = f(x, u_a(x))$.

        Since the linear policy $u_b$ is exponentially stabilizing within the region
        $S_a$,
        we similarly know that there must exist a bound $\tau_b \in \mathbb{R}_{>0}$ such that
        if $x(0) \in S_a \cap R_b$  
        then $x(t) \in S_b$ in finite time $t \leq \tau_b$ for the execution $\dot{x} = f(x, u_b(x))$.
        
        With $\bt_1 := (u_a, r_a)$ and $\bt_2 := (u_b, r_b)$, we then define and label the BT in Fig. \ref{fig:pendulum-bt} with
        \begin{equation}\label{eq:pendulum-bt}
            \bt_0 := Seq\left[\bt_1, \bt_2\right]
        \end{equation}
        The operating regions of (\ref{eq:pendulum-bt}) are then computed as
        \begin{equation}
            \Omega_0 = \mathbb{R}^n = \Omega_1 \cup \Omega_2, \quad \Omega_1 = R_1, \quad \Omega_2 = S_1 \cap (R_2 \cup S_2).
        \end{equation}
        Therefore, the execution (\ref{eq:bt_execution}) is computed with
        \begin{equation}\label{eq:pendulum-bt-control}
            u_0\left(x\right) = \begin{cases}
                u_1\left(x\right) & \text{if } x \in \Omega_1 \\
                u_2\left(x\right) & \text{if } x \in \Omega_2
            \end{cases}.
        \end{equation}
        
        Based on the stability analysis above, we then have the following in the context of Theorem \ref{theorem:convergence}:
        \begin{equation}\label{eq:pendulum-bt-constraint}
        \begin{aligned}
            \Lambda_1 &= \Omega_1 \cup \Omega_2 = \mathbb{R}^n \\
            \Lambda_2 &= \Omega_2
        \end{aligned}
         \qquad \qquad
        \begin{aligned}
            t' &= \tau_1 + \tau_2 \\
            N &= 2
        \end{aligned},
        \end{equation}
        with $L := \{1, 2\}$ and $\leq_f = \leq_S$, where $N=2$ because $L$ is totally ordered by $\leq_S$.
        The conclusion from (\ref{eq:pendulum-bt-constraint}) is that, for the execution (\ref{eq:bt_execution}),
        if $x(0) \in \Omega_1$ then $x(t) \in S_0$, where $S_0 = S_1 \cap S_2$, in finite time $t \leq t'$ within $N$ transitions.

\section{Conclusions} 
\label{sec:conclusion}
In this letter, we have formulated BTs in continuous-time and shown how they fit the formalism of a DDS
and the conditions under which solutions to their execution exist and are unique.
To do this, we embedded the order of the BT structure itself into the formulation.
These contributions allow the application of the rich literature
in hybrid dynamical systems 
\cite{branicky1998multiple, decarlo2000perspectives, hespanha1999stability}
to BTs in general.
Finally, we have provided
the conditions under which a general BT will be convergent to a goal.

\section*{Acknowledgment}
This  work  was  supported  by SSF  through  the  Swedish  Maritime Robotics Centre (SMaRC) (IRC15-0046).
    
{
\bibliographystyle{IEEEtran}
\bibliography{root.bib}}

\end{document}